# petBrain: A New Pipeline for Amyloid, Tau Tangles and Neurodegeneration Quantification Using PET and MRI

Pierrick Coupé[1], Boris Mansencal[1], Floréal Morandat[1], Sergio Morell-Ortega[2], Nicolas Villain[3,4], Jose V. Manjón[2], Vincent Planche[5,6]

1 CNRS, Univ. Bordeaux, Bordeaux INP, LABRI, UMR5800, F-33400 Talence, France

2 ITACA, Universitat Politècnica de València, 46022 Valencia, Spain

3 AP-HP Sorbonne Université, Hôpital Pitié-Salpêtrière, Department of Neurology, Institute of Memory and Alzheimer's Disease, 75013 Paris, France

4 Institut du Cerveau - ICM, Sorbonne Université, INSERM U1127, CNRS 7225, 75591 Paris, France

5 CHU Bordeaux, Service de Neurologie des Maladies Neurodégénératives, Centre Mémoire Ressources Recherche, F-33000 Bordeaux, France

6 Univ. de Bordeaux, CNRS, UMR 5293, Institut des Maladies Neurodégénératives, F-33000 Bordeaux, France

Corresponding author: Dr. Pierrick Coupé PhD., LaBRI - UMR 5800, 351 cours de la Libération, F-33405 Talence cedex, France ; Mail : pierrick.coupe@u-bordeaux.fr; Phone : +33 5 40 00 35 38

## Structured Abstract

**INTRODUCTION**: Quantification of amyloid plaques (A), neurofibrillary tangles ($T_2$), and neurodegeneration (N) using PET and MRI is critical for Alzheimer's disease (AD) diagnosis and prognosis. Existing pipelines face limitations regarding processing time, tracer variability handling, and multimodal integration.

**METHODS**: We developed petBrain, a novel end-to-end processing pipeline for amyloid-PET, tau-PET, and structural MRI. It leverages deep learning-based segmentation, standardized biomarker quantification (Centiloid, CenTauR, HAVAs), and simultaneous estimation of A, $T_2$, and N biomarkers. It is implemented in a web-based format, requiring no local computational infrastructure and software usage knowledge.

**RESULTS**: petBrain provides reliable, rapid quantification with results comparable to existing pipelines for A and $T_2$, showing strong concordance with data processed in ADNI databases. The staging and quantification of A/$T_2$/N by petBrain demonstrated good agreements with CSF/plasma biomarkers, clinical status and cognitive performance.

**DISCUSSION**: petBrain represents a powerful open platform for standardized AD biomarker analysis, facilitating clinical research applications.

# 1. Background

Alzheimer's disease (AD) is pathologically characterized by amyloid-β (Aβ) plaques, tau neurofibrillary tangles, and neurodegeneration. Recent recommendations from the Alzheimer's Association (AA) establish amyloid-PET as the gold standard for identifying brain amyloidosis (A) *in vivo*, and tau PET for quantifying and staging tauopathy ($T_2$). Neurodegeneration (N) can be defined using neuroimaging via structural MRI or FDG-PET.[1] With the recent development of disease-modifying therapies such as anti-amyloid monoclonal antibodies in early AD, there is an increasing demand for standardized, automated imaging biomarker quantification to identify eligible patients, monitor target engagement, and assess disease progression in clinical trials and real-world settings.[2]

Several challenges hinder the routine application of the biological diagnosis and staging of AD using the A/$T_2$/N imaging classification. First, manual or semi-automated workflows introduce variability, as different processing pipelines, regions of interest and thresholding techniques can yield inconsistent results.[3,4] Second, reproducibility is limited across research centers due to differences in scanner types, acquisition protocols, and tracer standardization. Moreover, time-consuming and labor-intensive procedures make it difficult to implement these biomarkers in large-scale clinical trials and multi-center academic studies. Finally, clinical translation is constrained, as biomarker quantification must be accurate, fast, and operator-independent for real-world applications.

In the literature, various software solutions have been developed to analyze individual biomarkers of the A/$T_2$/N biological framework. Several solutions exist for amyloid-PET imaging.[5,6] For instance, the Centiloid (CL) Project proposed a pipeline based on SPM8[7], while the ADNI consortium developed a pipeline based on FreeSurfer.[8] The CL scale[7] is now a widely adopted method, endorsed for AD diagnosis by the European Medicines Agency (EMA). The CL scale standardizes amyloid-PET data across tracers and cohorts, facilitating cross-study comparisons. Moreover, some amyloid "PET-only" pipelines that do not require associated structural MRI, are now available.[9,10] For tau-PET imaging, fewer solutions are available[11] but the CenTauRz (CTRz) Project recently introduced the CapAIBL pipeline, to define a universal standard approach similar to the one used in the CL project.[12] To quantify neurodegeneration using structural MRI, many generalist tools exist to segment brain structures, but few offer dedicated scores for AD. Most current methods focus on key structures such as the hippocampus to assist clinical interpretation of T1w MRI.[13,14] The Hippocampal-Amygdalo-Ventricular Atrophy score (HAVAs) is, for instance, a structural biomarker that quantifies atrophy in the hippocampus, amygdala, and ventricles using normative and pathological lifespan models, which has been validated as a sensitive marker of AD-related neurodegeneration.[13]

Simple and integrated technical solutions enabling the simultaneous analysis of all pathological components of AD in individual patients would facilitate the broader application of the A/$T_2$/N biological framework. Recently, the B-PIP pipeline has been proposed to produce A/$T_2$ biomarkers based on CL and Tau Standardized uptake value ratio (SUVr),[15] but it does not provide CTRz universal values for standardizing tau imaging studies. Furthermore, no user-friendly, end-to-end solutions are currently available that can simultaneously estimate A/$T_2$/N from neuroimaging data. To address these issues, we propose a fully automated A/$T_2$/N pipeline, integrating state-of-the-art neuroimaging biomarkers for amyloid-PET, tau-PET, and MRI-derived neurodegeneration measures. This pipeline, named petBrain, ensures standardization, reproducibility, and scalability, making it well-suited for clinical research applications. In our pipeline, amyloid load (A) is quantified using the Centiloid scale,[7] tau neurofibrillary tangles ($T_2$) are assessed using CTRz scale[12] and neurodegeneration (N) is measured with HAVAs.[13] In this article, we describe petBrain implementation, its methodological validation, and the assessment of its biological and clinical correlates, demonstrating its potential for biological staging of AD. To enable its simple, free, and large-scale use, the petBrain pipeline is freely available online on the user-friendly volBrain platform (www.volbrain.net).

# 2. Methods

## 2.1. Participants

### GAAIN Datasets

This study used datasets collected from the publicly available Global Alzheimer's Association Interactive Network (GAAIN) repository (https://www.gaain.org). First, we used 499 amyloid-PET images from the Centiloid project obtained with five different amyloid-PET tracers: $^{11}$C-PiB (PiB), $^{18}$F-Florbetapir (FBP), $^{18}$F-Flutemetamol (FTM), $^{18}$F-Florbetaben (FBB), and $^{18}$F-NAV4694 (NAV). These images were accompanied by corresponding 3D T1-w MRI scans from patients with AD clinical syndrome, frontotemporal dementia (FTD), and mild cognitive impairment (MCI), as well as young and elderly cognitively unimpaired controls (Table 1 and supplementary material section 1). PET acquisition times varied by tracer: 50 to 70 minutes post-injection for PiB and NAV, 50 to 60 minutes post-injection for FBP, and 90 to 110 minutes post-injection for FTM and FBB. Additionally, we used 100 tau PET images from the CTRz project, acquired using a single tau tracer, $^{18}$F-Flortaucipir (FTP), along with corresponding 3D T1-w MRI scans from patients with dementia and elderly cognitively unimpaired controls.

### ADNI Dataset

To validate our pipeline on an external dataset, we used the Alzheimer's Disease Neuroimaging Initiative (ADNI) database (https://adni.loni.usc.edu). ADNI is a comprehensive study aimed at developing and validating biomarkers for AD diagnosis and progression. It includes imaging data (MRI, PET), clinical assessments, and biological measurements from blood and cerebrospinal fluid (CSF), collected from participants with normal cognition, mild cognitive impairment, and dementia. In this study, we used 821 subjects having at the same time a T1w MRI, an amyloid-PET and a tau-PET in order to validate our A/$T_2$/N pipeline (Table 1). The amyloid PET tracers were FBB and FBP, while the tau PET tracers were FTP, $^{18}$F-MK6240 (MK) and $^{18}$F-PI2620 (PI). The participants were selected within the clinical labels

"Cognitively Normal (CN)", "MCI", and "AD dementia" provided by ADNI. Moreover, the amyloid status provided by the ADNI database has been used to classify subjects as amyloid-negative (A-) and amyloid-positive (A+). This status has been evaluated by ADNI with their own amyloid-PET pipeline.[8] Finally, we collected individual global cognitive performance (i.e., Clinical Dementia Rating-sum of boxes (CDR-sb), Mini Mental State Examination (MMSE), and Montreal Cognitive Assessment (MoCA)) as well as their CSF and plasma biomarkers concentrations when available.

Table 1 : **GAAIN-Development and ADNI-Validation samples with amyloid and tau PET tracers used to calibrate and validate petBrain**. The clinical labels yCN (young Cognitively Normal), CN (Cognitively Normal), MCI (Mild Cognitive Impairment) and Dementia were the labels provided by the GAAIN and ADNI datasets (see section 1 of supplementary material for more details).

| | **Amyloid Tracers per scan** | **Tau Tracers per scan** | **Age** | **Participant** |
|---|---|---|---|---|
| **GAAIN N = 375** | FBB / FBP / PiB / FTM / NAV | FTP / MK / PI | Mean (SD) | N |
| **yCN** | 10 / 13 / 91 / 24 / 10 | N/A | 32.3 ± 7.4 | 77 |
| **CN** | 6 / 6 / 47 / 10 / 25 | 50 / 0 / 0 | 67.6 ± 8.4 | 97 |
| **MCI** | 9 / 10 / 49 / 20 / 10 | N/A | 74.4 ± 7.1 | 49 |
| **Dementia*** | 10 / 17 / 102 / 20 / 10 | 50 / 0 / 0 | 68.1 ± 8.5 | 152 |
| **ADNI N = 831** | | | | |
| **CN A-** | 146 / 188 / 0 / 0 / 0 | 320 / 5 / 9 | 69.6 ± 7.1 | 344 |
| **CN A+** | 51 / 94 / 0 / 0 / 0 | 142 / 0 / 3 | 73.6 ± 7.8 | 145 |
| **MCI A-** | 63 / 65 / 0 / 0 / 0 | 123 / 3 / 2 | 71.7 ± 8.1 | 128 |
| **MCI A+** | 58 / 78 / 0 / 0 / 0 | 126 / 8 / 2 | 74.8 ± 7.2 | 136 |
| **Dementia** A-** | 10 / 4 / 0 / 0 / 0 | 11 / 2 / 1 | 74.0 ± 8.9 | 14 |
| **Dementia** A+** | 26 / 38 / 0 / 0 / 0 | 58 / 5 / 1 | 76.4 ± 9.5 | 64 |

[*] Dementia diagnoses including AD clinical syndromes and FTD.
[**] Dementia with AD clinical syndrome.

## 2.2. The petBrain pipeline

### Related works

The CL scale, designed to standardize amyloid PET quantification across different radiotracers,[7] is computed through a pipeline that involves co-registration of MRI and PET images followed by spatial normalization to Montreal Neurological Institute (MNI) space using

SPM8.[16] After transformation to this common space, standardized anatomical masks are applied, including a predefined global cortical target region—encompassing the prefrontal, parietal, temporal cortices and the frontal gyrus—as well as a reference region-of-interest (ROI), typically the whole cerebellum. A key strength of this approach lies in its open-source availability for academic use. However, the method presents several limitations: it requires manual reorientation of both PET and MRI volumes, access to a licensed MATLAB environment, and specific technical expertise in SPM8. In addition, the use of static anatomical masks may limit its adaptability and potentially reduce the accuracy of SUVr estimates in the presence of inter-individual anatomical variability.

A comparable framework has been developed for tau PET imaging by the CTRz Project, which offers both a traditional implementation using SPM8, and a cloud-based version accessible through the CapAIBL platform (https://capaibl-milxcloud.csiro.au).[12] The SPM8-based pipeline shares the advantages and limitations of the original CL method, offering open academic access but requiring local computational resources and expertise in neuroimaging. In contrast, CapAIBL delivers a user-friendly, web-based solution that removes the need for software installation or advanced technical skills. In line with the standards of the volBrain platform, CapAIBL ensures compliance with medical data privacy regulations by mandating anonymized image submissions. Moreover, as with the volBrain platform, all uploaded data are automatically and permanently deleted from secure servers after a fixed retention period, unless users have explicitly provided consent for their reuse in research.

To address the limitations of fixed region-of-interest (ROI) masks, the ADNI pipeline[8] introduced an alternative anatomical delineation approach using subject-specific ROIs derived from FreeSurfer-based segmentation[17] of structural MRI, coupled with SPM12 registration. This pipeline offers improved anatomical precision and relies entirely on open-access software. However, it also presents notable drawbacks: the FreeSurfer segmentation process is computationally intensive (typically >15 hours per subject) and requires substantial expertise for installation and operation.

Until recently, most existing pipelines have focused on measuring either A or $T_2$ biomarkers individually. A recent advancement is the B-PIP pipeline[15], an updated version of the ADNI approach, which supports simultaneous estimation of both A and $T_2$ biomarkers using FreeSurfer and SPM12. This marks a significant step toward implementing the A/$T_2$/N framework in clinical research. However, this pipeline still relies on SUVr quantification for Tau, which limits its robustness across different tau tracers, and currently lacks full integration of the N biomarker.

To leverage the strengths of existing methods while addressing their limitations, we have developed a novel end-to-end pipeline with the following key features:

- Subject-specific ROI masks generated in under 15 minutes via deep learning-based segmentation of 3D MRI, significantly reducing processing time compared to FreeSurfer.
- Robust biomarker quantification using standardized CL and CTRz scales, enabling consistent analysis across different amyloid and tau tracers (unlike SUVr-based approaches).
- Simultaneous estimation of A, $T_2$, and N biomarkers within a unified processing framework.
- Web-based implementation requiring no local computational infrastructure or expertise in image processing software.

## Global overview

An overview of the proposed petBrain pipeline is presented in Figure 1. The pipeline delivers a comprehensive quantification of AD biomarkers by providing: (i) global amyloid and tau burden expressed on standardized scales – CL and CTRz, respectively; (ii) the degree of neurodegeneration estimated via the Hippocampal-Amygdalo-Ventricular Atrophy score (HAVAs); and (iii) the corresponding binary pathological status for each biomarker (A+/A−, $T_2$+/$T_2$−, N+/N−). Thresholds for classification are derived from externally validated studies:

AMYPAD for CL, the CenTauR project for CTRz,[12,18] and the original HAVAs publication for neurodegeneration.[13]

Additionally, petBrain provides automated segmentation and volumetric quantification of 132 brain structures using the AssemblyNet framework.[19] Leveraging a normative lifespan model, the pipeline also estimates deviations from typical aging trajectories for each structure.[20] Neurodegenerative status (N+/N−) is determined using the HAVAs score, which is derived from a pathological lifespan model based on a composite of amygdala, hippocampus, and inferior lateral ventricle volumes – a combination shown to be highly sensitive to typical Alzheimer's-related atrophy.[13,21] Furthermore, petBrain computes SUVr for both amyloid and tau-PET across 122 gray matter regions, corresponding to the 132 segmented structures excluding white matter and cerebrospinal fluid (CSF) regions.

To ensure broad accessibility and eliminate the need for local computational resources or specialized personnel, the petBrain pipeline is made available through the open-access volBrain platform (www.volbrain.net). Launched in 2015, volBrain offers a suite of fully automated pipelines for brain MRI segmentation and computer-aided diagnosis. To date, the platform has processed over 700,000 images for more than 13,000 users worldwide (www.volbrain.net/users). As with all volBrain pipelines, petBrain outputs a comprehensive results package, including a PDF report and corresponding CSV files containing all quantitative measurements. Additionally, users can download the full set of processed 3D images and segmentations as a compressed archive of NIfTI files, facilitating further visualization or analysis.

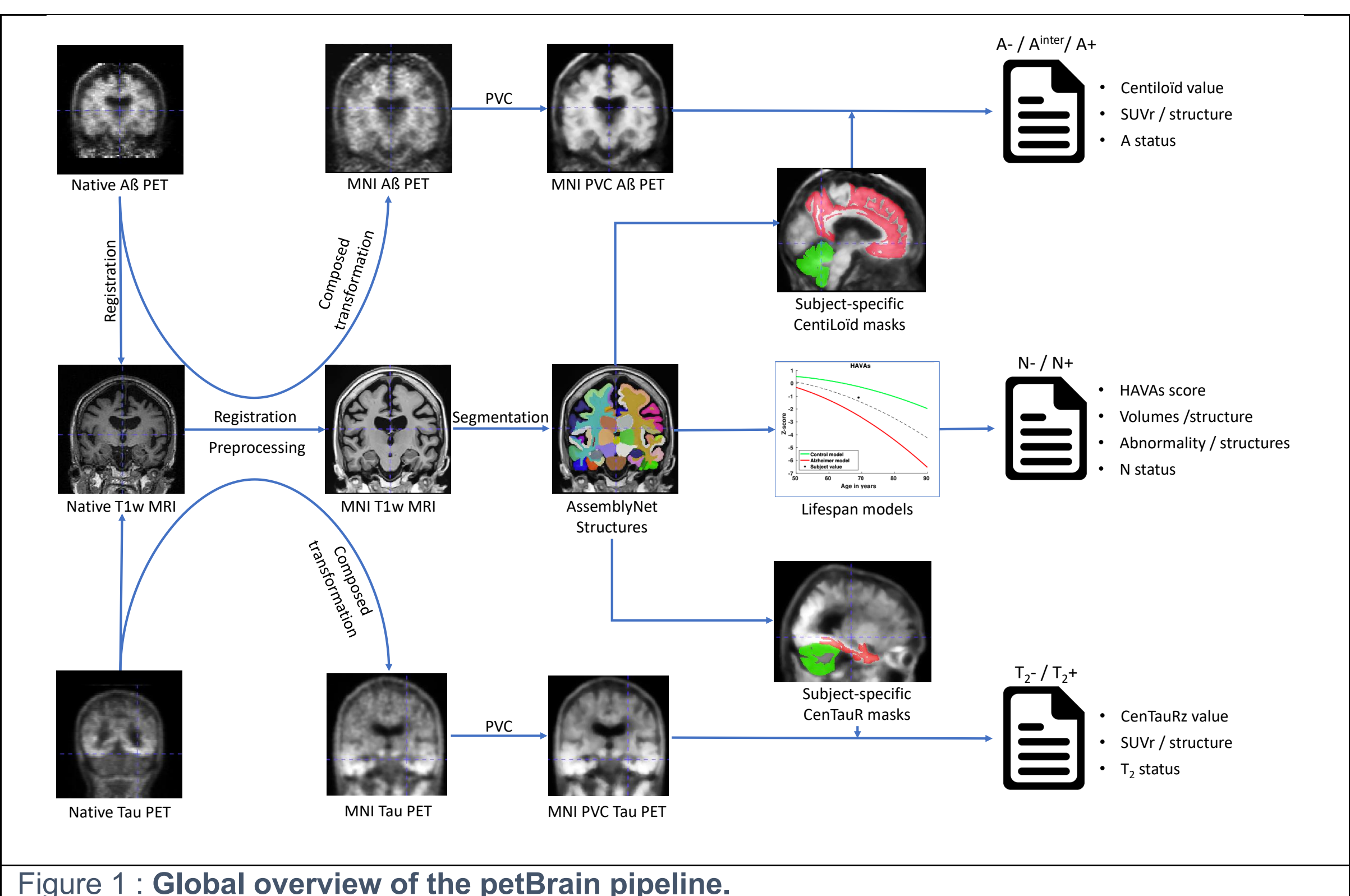

Figure 1 : **Global overview of the petBrain pipeline.**

## T1w MRI processing

The native T1w MRI of the subject is processed through the AssemblyNet pipeline.[19] This pipeline includes a preprocessing step to harmonize data followed by a segmentation step based on deep learning.

**Preprocessing**: The preprocessing step is based on the following steps: denoising[22], ii) inhomogeneity correction,[23] affine registration into the MNI space (181 x 217 x 181 voxels at 1 x 1 x 1 $mm^3$),[24] tissue-based intensity normalization[25] and intracranial extraction.[26] Finally, image intensities are z-scored (i.e., centralized and normalized) within the brain mask and the background voxels were set to zero.

**Volume values**: The segmentation step is based on a large ensemble of deep learning models in order to provide accurate and robust segmentation of 132 structures as described in the Neuromorphometrics protocol (https://neuromorphometrics.com/Seg/). Finally, the volume values (native and normalized by intracranial volume) are estimated for all the structures. The normalized volume values are compared to a normal aging model[20] to detect abnormalities indicated in the final PDF report.

**N status**: To estimate the neurodegenerative status of the subject under study, the HAVAs score is estimated using the AssemblyNet segmentation of the hippocampus, the amygdala and the inferior lateral ventricle. Based on lifespan modeling of normal aging and aging with AD[21], the threshold between N+ and N- is adapted according to the subject's age as described in the original article (i.e., a probability higher than 0.5 is considered as N+).[13] We ensured that ADNI subjects involved during HAVAs model construction were not included in this study.

### PET scan processing

**Preprocessing**: The native PET scan are first rigidly registered to the native T1w MRI.[24] By composing the obtained transformation with the transformation obtained from the registration of the T1w MRI into the MNI space, the PET was directly interpolated into the MNI space using a single interpolation to avoid additional blurring. Afterwards, a super-resolution method involving self-similarity within PET, and image priors based on T1w MRI, is used to compensate for partial volume effect of the PET scan.[27]

**SUVr values**: The partial volume corrected (PVC) PET is then converted into SUVr using the AssemblyNet segmentation. As proposed by the CL and CTRz projects, we use the whole cerebellum for amyloid-PET and cerebellar gray matter for tau-PET as the reference ROI. Finally, SUVr values are estimated for all subcortical and cortical structures provided by AssemblyNet for amyloid and tau-PET and provided as CSV files.

**Centiloid/CenTauRz values:** The aim of the CL and CTRz projects was to harmonize the estimation of amyloid and tau load, and to homogenously define A+/A- or $T_2$+/$T_2$- status across tracers, scanners and pipelines. To this end, both projects proposed predefined masks in the MNI space (the same mask whatever the subject's anatomy). To take advantage of having a subject-specific structure segmentation obtained from AssemblyNet, we defined a list of structures as CL and CTRz cortical masks and used the same reference masks as for SUVr estimation. For the estimation of our Centiloid mask, as suggested by Klunk et al.[7], the GAAIN-PiB dataset was used to establish the list of the most discriminate structures between young

CN (yCN) A- subjects and A+ patients with dementia (see supplementary Table 1 for the list of selected structures). For our CTRz mask, we used the *a priori* list of structures validated in several studies[28,29] – entorhinal area, amygdala, parahippocampal gyrus, fusiform gyrus, inferior and middle temporal gyrus, and temporal pole (see supplementary Table 2 for the list of selected structures). As subsequent results demonstrate, this mask closely resembles the MetaTemporal mask proposed by the CTRz project and B-PIP pipelines.

**A and $T_2$ status**: The CL and CTRz values are then used to estimate the A status and $T_2$ status of each subject. During the experiments presented in this article, we used the common threshold of 24.1 CL for Amyloid.[30] However, for the online version of the pipeline, we decided to follow the recommendation of AMYPAD[18] with CL <10 considered as A- , CL between 10 and 30 as intermediate $A^{inter}$, and CL>30 as A+. For CTRz, we used the threshold proposed by the CenTaur project with a CTRz < 2 as $T_2$- and $T_2$+ otherwise.[12]

Overall, the total processing time of the petBrain pipeline is approximately 20 minutes: ~15 minutes for global T1-weighted MRI preprocessing and segmentation, ~1 minute for each rigid PET-to-T1 registration, and ~1 minute per PVC step. This runtime is comparable to that of SPM-based pipelines and substantially faster than those based on FreeSurfer.

# 3. Results

## 3.1. Centiloid Calibration

### PiB calibration

First, the GAAIN-PiB dataset was used to convert the original SPM8 Centiloid pipeline and petBrain. To this end, we performed the Level-1 calibration procedure[7] and we obtained the following equation:

$$CL = 100 \times ({}^{PiB}SUVr_{petBrain} - 0.9659) / (1.8972 - 0.9659) \qquad \text{Eq. (1)}$$

Therefore, an individual SUVr PiB value ($^{PiB}SUVr_{petBrain}$) obtained using our petBrain pipeline can be converted into CL value.

Second, once calibrated, we compared the CL value obtained by petBrain with the CL values published by the Centiloid Project[7] using the official SPM8-based pipeline to ensure that our Level-1 calibration meets Centiloid method criteria. As shown in Supplementary Fig. 1, our calibration yielded to very high correlation between CL PiB published by the Centiloid Project[7] and CL PiB obtained with petBrain. Moreover, the linear regression parameters (y=0.99x + 0.57; $R^{2=}$0.99) fitted in the Centiloid Project criteria (*i.e.,* a slope between 0.98 and 1.02, an intercept between -2 and +2 CL, and an $R^2$ > 0.98).

### Amyloid tracers' calibration

Once Level-1 calibration was obtained for PiB tracer, we estimated the Level-2 calibration for all other considered amyloid tracers (i.e., FBP, FBB, FTM and NAV). First, linear regressions were estimated between PiB and other tracers using the corresponding Centiloid Project datasets: the obtained regressions $R^2$ are summarized in Table 2 (see also supplementary Figure 2 for more details). Afterwards, the conversion between amyloid tracers and CL scale was obtained using the equation provided in Klunk et al,[7]. The resulting conversion equations are also presented in Table 2.

Table 2: **Equations used for Centiloid calibration of petBrain.**

| Amyloid Tracers | $R^2$ | Conversion equations |
| --- | --- | --- |
| **PiB** | N/A | $CL = 107.3768 \times {}^{PiB}SUVr_{petBrain} - 103.7152$ |
| **FBP** | 0.92 | $CL = 194.8721 \times {}^{FBP}SUVr_{petBrain} - 191.8315$ |
| **FBB** | 0.97 | $CL = 165.2828 \times {}^{FBB}SUVr_{petBrain} - 158.0409$ |
| **FTM** | 0.95 | $CL = 141.1563 \times {}^{FTM}SUVr_{petBrain} - 128.3451$ |
| **NAV** | 0.99 | $CL = 104.8498 \times {}^{NAV}SUVr_{petBrain} - 102.6445$ |

## 3.2. CenTauR calibration

The CenTauR Project provides several predefined masks (i.e., universal, mesial-temporal, meta-temporal, temporo-parietal and frontal).[12] Our first step was to find the CenTauR's mask that was the most correlated with our subject-specific mask. To this end, we estimated the relationships between published $^{FTP}SUVr_{CTR}$ obtained for each mask and the $^{FTP}SUVr_{petBrain}$ estimated with petBrain on the FTP dataset. Supplementary Table 3 presents the $R^2$ obtained through linear regression. These results show that the predefined Meta-temporal CenTauR mask is the most similar to our subject-specific mask, including the entorhinal area, amygdala, parahippocampal gyrus, fusiform gyrus, inferior and middle temporal gyrus, and temporal pole (see section 5 of the supplementary material for the list of selected structures). Consequently, we used this mask in the following steps.

Second, we used the linear regression between the published $^{FTP}SUVr_{CTR}$ by the CenTauR Project using SPM8-based pipeline over the Meta-temporal mask and the $^{FTP}SURv_{petBrain}$ estimated with petBrain. This yielded the following Level-1 calibration equation:

$$^{FTP}SUVr_{CTR} = (^{FTP}SUVr_{petBrain} - 0.2222) / 0.7646 \qquad \text{Eq. (2)}$$

Third, once calibrated, we compared the $CTR_z$ values obtained by petBrain with the $CTR_z$ values published by the CenTauR Project[12] using their SPM8 pipeline. Supplementary Figure 3 presents the results of this comparison using their Meta-temporal mask ($y=0.9804x +0.096$; $R^2=0.9803$). As for CL, the obtained linear regression parameters are a slope between 0.98 and 1.02 and an $R^2 > 0.98$.

Contrary to the Centiloid project, where a dataset is available for each amyloid tracer, the CenTauR project provides only one dataset for the FTP tracer. Therefore, we directly used the pipeline calibration obtained on FTP (see Eq. 2) to adapt petBrain to other tau tracers. By using the equations provided in Villemagne et al.[12] for the Meta-Temporal mask, we obtained the conversion equations presented in Table 3 for five other tau-tracers: $^{18}$F-RO948 (RO), $^{18}$F-MK-6240 (MK), $^{18}$F-GTP1 (GTP), $^{18}$F-PM-PBB3 (PBB3) and $^{18}$F-PI2620 (PI).

Table 3 : **Equations used for CenTauR calibration of petBrain.**

| Tau Tracer | $R^2$ | Conversion equations |
|---|---|---|
| **FTP** | 0.98 | CTRz = 16.9370 x ${}^{FTP}SUVr_{petBrain}$ - 19.1334 |
| **RO** | N/A | CTRz = 17.2116 x ${}^{RO}SUVr_{petBrain}$ - 20.0144 |
| **MK** | N/A | CTRz = 12.2417 x ${}^{MK}SUVr_{petBrain}$ - 12.7801 |
| **GTP** | N/A | CTRz = 12.5556 x ${}^{GTP}SUVr_{petBrain}$ - 13.8899 |
| **PBB3** | N/A | CTRz = 15.4067 x ${}^{PBB}SUVr_{petBrain}$ - 14.6334 |
| **PI** | N/A | CTRz = 10.1753 x ${}^{PI}SUVr_{petBrain}$ - 11.5909 |

## 3.3. Validation of petBrain on ADNI

### Comparison of amyloid and tau loads with state-of-the-art pipelines

To further validate our pipeline, we conducted a comparative analysis of amyloid and tau burden estimates using an independent dataset. Specifically, we compared values derived from our petBrain pipeline with those obtained using SPM-based Centiloid and CenTauR pipelines, as well as the B-PIP pipeline.[15] For the SPM-based methods, we employed the last version of SPM (i.e., SPM25). [31] Centiloid values were computed using the standard cortical target mask and the whole cerebellum as the reference region, consistent with the original calibration protocol. For CenTauR quantification, we employed the MetaTemporal target mask and the cerebellar grey matter as the reference region, in order to ensure maximal consistency with the masks used in the B-PIP and petBrain pipelines. Finally, B-PIP values correspond to the scalar scores provided directly by the ADNI consortium.

First, we compared SPM-based Centiloid pipelines and petBrain using linear regression and intraclass correlation coefficient (ICC). As shown in Figure 2 (left panel), the two pipelines yielded highly correlated results ($R^2$ = 0.94; ICC = 0.97). Second, we compared CTRz values produced by SPM-based CenTauR pipelines and petBrain. Figure 2 (right panel) also demonstrates a good correlation between pipelines ($R^2$ = 0.84; ICC = 0.92).

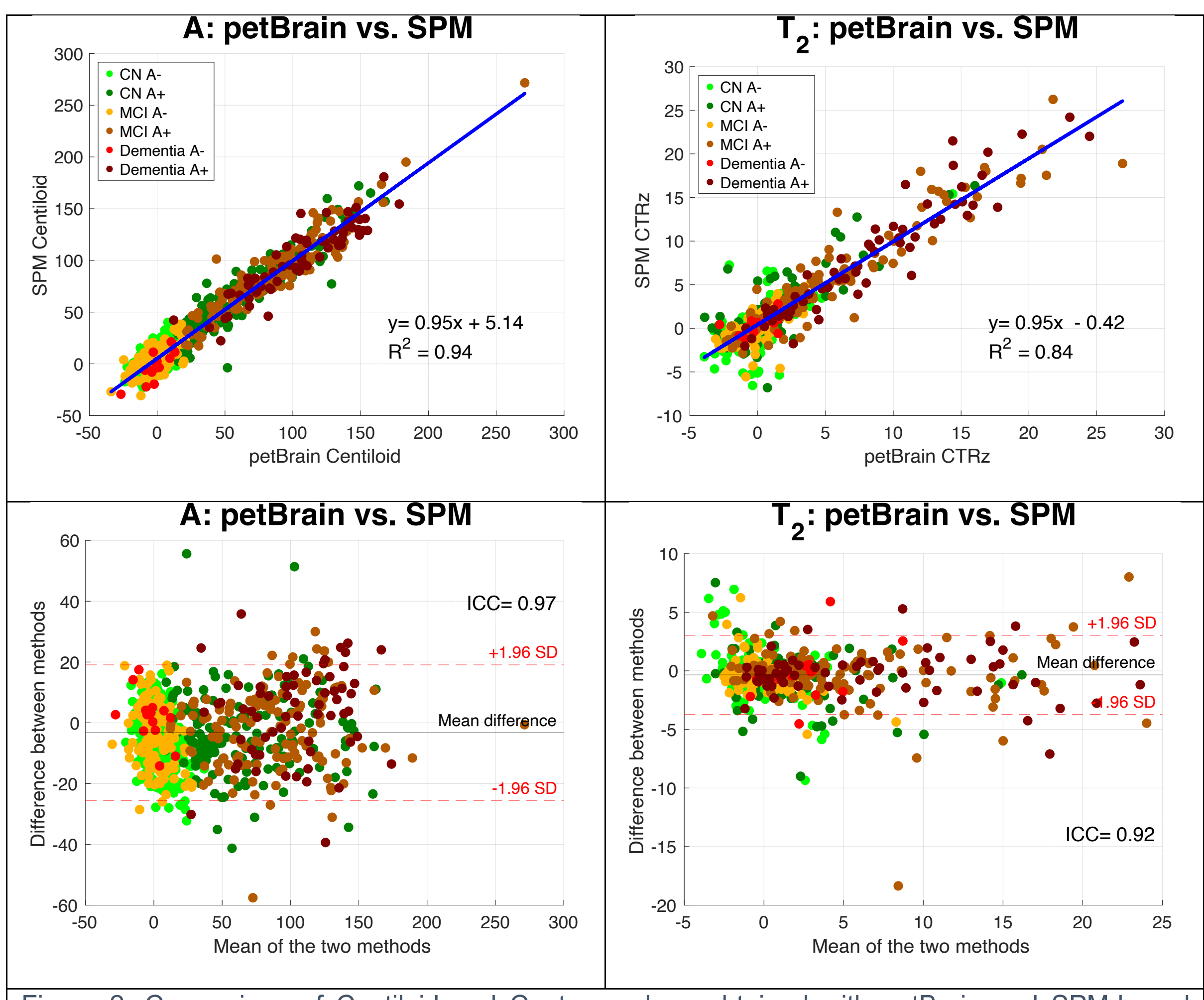

Figure 2: Comparison of Centiloid and Centaur values obtained with petBrain and SPM-based pipelines.

Afterwards, we assessed the concordance between B-PIP and petBrain CL values. As shown in Figure 3 (left panel), the two pipelines yielded highly correlated results ($R^2$ = 0.96; ICC = 0.98). Notably, a subset of CN A+ individuals exhibited systematically higher CL values with petBrain compared to B-PIP, as illustrated in the Bland-Altman plot. Visual quality control of these outliers, using images and automated PDF reports generated by petBrain, revealed no apparent processing errors. Such verification was not feasible for B-PIP due to the absence of accessible intermediate outputs. The Bland–Altman analysis further revealed a slight but consistent upward bias for higher CL values, suggesting a non-uniform deviation between pipelines at the upper quantification range. We then compared Tau PET quantification between pipelines by evaluating Tau standardized SUVr in MetaTemporal mask, as B-PIP does not generate CTRz metrics. Figure 3 (right panel) demonstrates a similarly strong correlation between pipelines for Tau SUVr ($R^2$ = 0.95; ICC = 0.96), though greater discrepancies were observed at higher SUVr values.

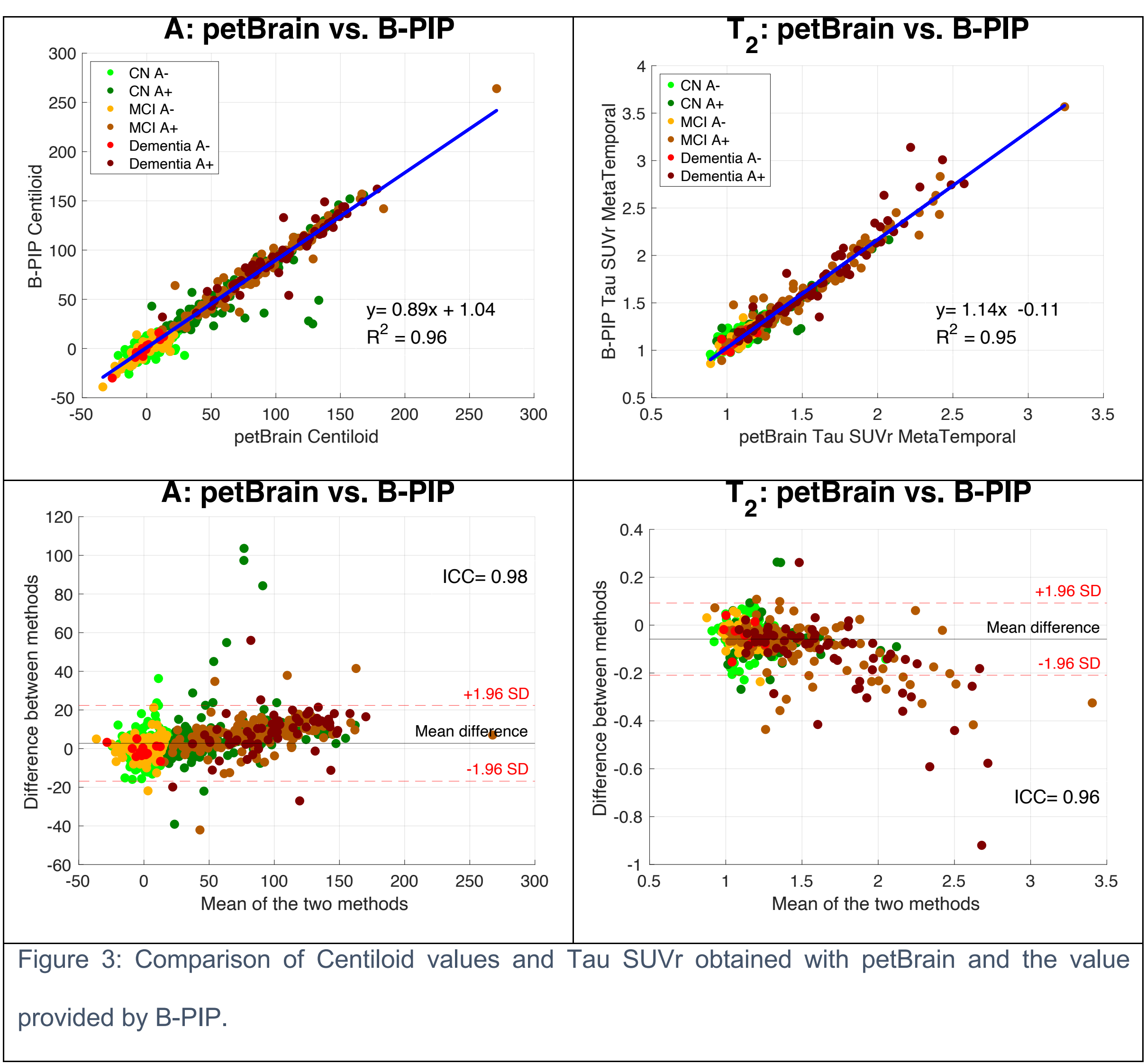

Figure 3: Comparison of Centiloid values and Tau SUVr obtained with petBrain and the value provided by B-PIP.

Finally, we compared SPM-based pipelines and B-PIP. As shown in Figure 4 (left panel), Centiloid values obtained from the two pipelines were highly correlated ($R^2$ = 0.90; ICC = 0.95). Similar to observations made with petBrain, a subset of CN A+ individuals showed systematically higher CL values with the SPM-based pipeline compared to B-PIP, suggesting possible failure in the B-PIP processing for these cases. We then compared Tau PET quantification using SUVr values from the SPM-based CenTauR pipeline versus B-PIP. As shown in Figure 4 (right panel), this comparison also demonstrated a strong correlation ($R^2$ =

0.84; ICC = 0.91), although greater dispersion — particularly among CN subjects — was observed.

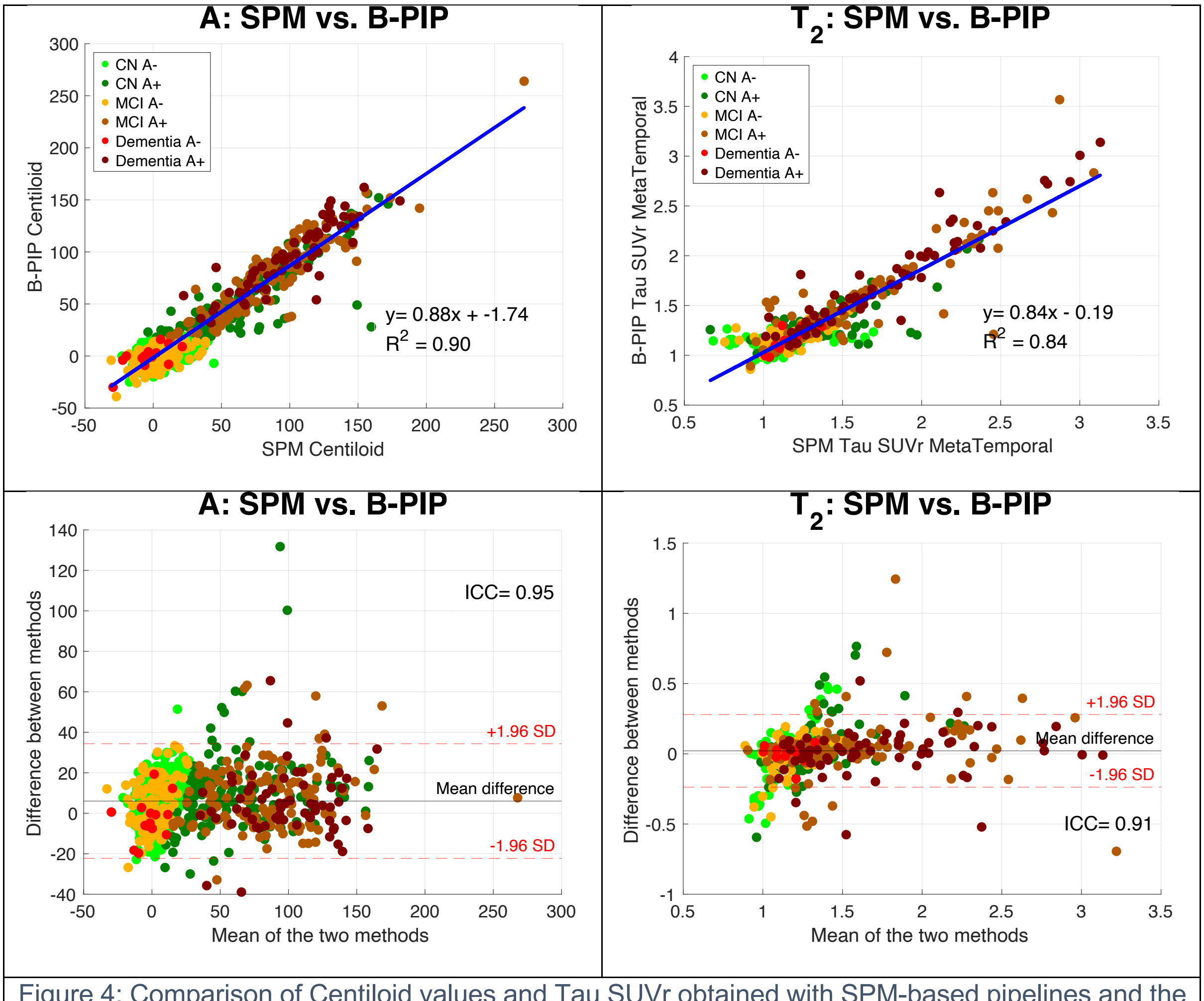

Figure 4: Comparison of Centiloid values and Tau SUVr obtained with SPM-based pipelines and the value provided by B-PIP.

Together, these findings highlight a high level of concordance across the three pipelines. Notably, the strongest correlation was observed between the two pipelines employing subject-specific anatomical masks (petBrain and B-PIP), particularly for Tau SUVr quantification ($R^2$=0.95, compared to $R^2$=0.84 for both petBrain vs. SPM and SPM vs. B-PIP). However, fewer extreme outliers were observed between petBrain and the SPM-based pipelines than between B-PIP and SPM-based pipelines, suggesting better robustness of petBrain and SPM.

### Correlations with CSF and plasma biomarkers

We subsequently assessed the correspondence between CL values, meta-temporal tau-PET SUVr, CTRz, HAVAs and AD fluid biomarkers. To this end, we fitted linear mixed-effects models (adjusted for age, sex, and APOEε4 status) to evaluate the associations between fluid biomarkers and values derived from SPM-based pipelines, B-PIP and petBrain. Table 4 summarizes the statistical results. Three pipelines demonstrated comparable predictive performance across fluid biomarkers, with no significant differences observed in the explained variance ($R^2$) based on Steiger's Z-test.

Table 4: **Association of Centiloid (CL), tau SUVr, CenTauRz (CTRz) and the Hippocampo-Amygdalo-Ventricular Atrophy score (HAVAs) produced by petBrain, B-PIP and SPM-based pipelines with AD fluid biomarkers.** The linear mixed models were adjusted for age, sex, and APOEε4. The used metrics were p-value of the F-test and $R^2$.

| | p-value | $R^2$ |
|---|---|---|
| **CL ~ CSF Aβ42/40 (N=353)** | | |
| **petBrain** | 2.46e-58 | 0.546 |
| **B-PIP** | 1.02e-57 | 0.542 |
| **SPM-Centiloid** | 3.3e-56 | 0.532 |
| **CL ~ CSF p-tau (N=350)** | | |
| **petBrain** | 7.1e-27 | 0.311 |
| **B-PIP** | 2.18e-25 | 0.297 |
| **SPM-Centiloid** | 2.15e-24 | 0.287 |
| **CL ~ log(Plasma p-tau217) (N=397)** | | |
| **petBrain** | 2.49e-65 | 0.542 |
| **B-PIP** | 2.71e-63 | 0.531 |
| **SPM-Centiloid** | 2.3e-61 | 0.520 |

| Meta-temporal tau SUVr ~ log(Plasma p-tau217) (N=397) | | |
|---|---|---|
| **petBrain** | 8.23e-54 | 0.475 |
| **B-PIP** | 8.24e-55 | 0.482 |
| **SPM-Centaur** | 1.06e-54 | 0.481 |
| **CTRz ~ log(Plasma pau217) (N=397)** | | |
| **petBrain** | 1.48e-56 | 0.492 |
| **SPM-Centaur** | 1.19e-55 | 0.487 |
| **HAVAs ~ log(Plasma NfL) (N=398)** | | |
| **petBrain** | 2.65e-44 | 0.413 |

### Comparison of A/T$_2$/N staging with clinical staging

To further assess the performance of the petBrain pipeline, we examined the automated petBrain-derived A/T$_2$/N staging across CN individuals and patients with MCI or dementia. Figure 5 displays the results for 821 subjects from the ADNI cohort.

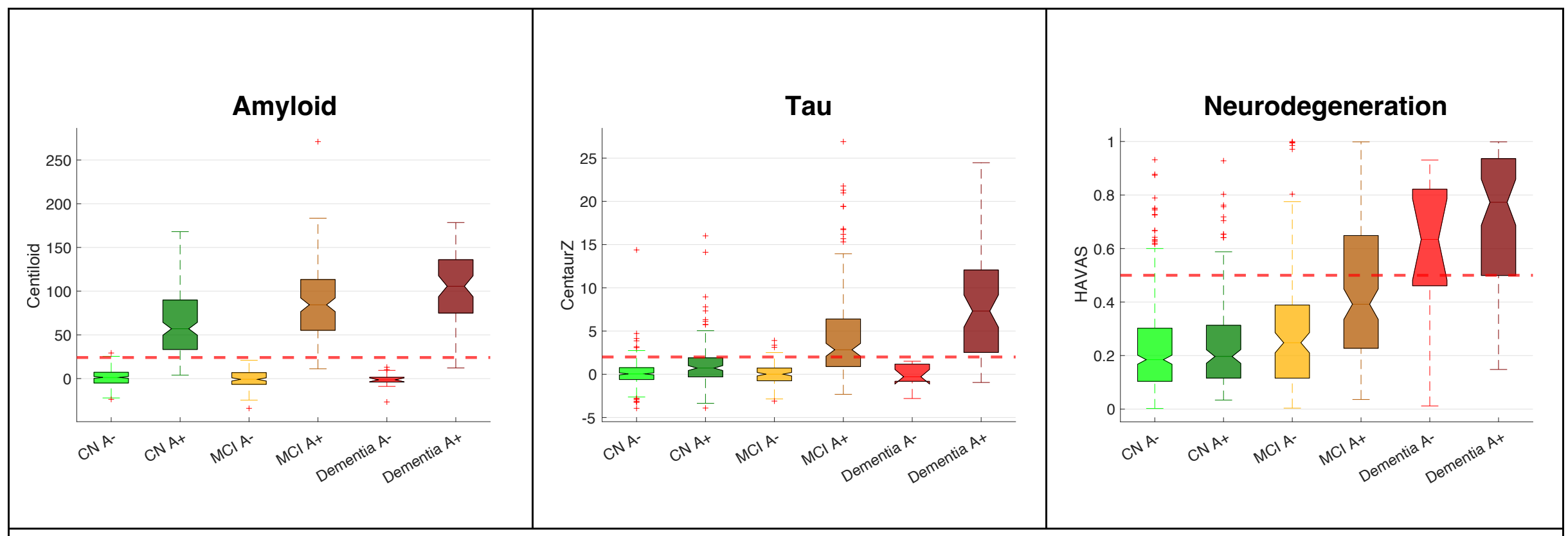

Figure 5 : Centiloid (Amyloid), CenTauRz (Tau) and HAVAs (Neurodegeneration) scores on the ADNI dataset (N=821) using petBrain. Amyloid status was established with the B-PIP PET pipeline and ADNI thresholds. The horizontal lines indicate the pathological threshold for each biomarker.

To investigate amyloid load across clinical stages in A+ individuals, we performed a one-way analysis of variance (ANOVA) comparing amyloid-positive CN subjects, and patients with MCI or dementia, followed by Tukey's test for *post-hoc* multiple comparisons. The analysis revealed a strongly significant association between amyloid load obtained from petBrain CL and clinical staging (mean CL value in A+ CN= 64.73; A+ MCI = 84.23; A+ dementia = 104.35; $p < 1\times10^{-10}$). We also found a strong association between tau load or neurodegeneration measured with petBrain and clinical staging in A+ participants ($T_2$: mean CTRz value in A+ CN= 1.09; A+ MCI = 4.62; A+ dementia = 7.88; $p < 1\times10^{-19}$. N: mean HAVAs probability in A+ CN= 0.25; A+ MCI = 0.44; A+ dementia = 0.69; $p < 1\times10^{-31}$). *Post-hoc* comparisons confirmed that each clinical A+ group differed significantly from the others in terms of A, $T_2$, and N measurements.

### Correlations of A/$T_2$/N quantification with cognitive scores

Finally, we evaluated the relationship between global cognitive performance and petBrain-derived biomarkers, quantified with the CL scale (A), CTRz ($T_2$) and HAVAs (N). Linear mixed-effects models were used to predict CDR-sb, MMSE, and MoCA scores, adjusting for age, sex, APOEε4 status, and education level. The results are presented in Table 5. The combined quantification of A/$T_2$/N explained a greater proportion of the variance ($R^2$) in cognitive measures than amyloid load (CLs) alone across all cognitive scales (adjusted Steiger's test on $R^2$). This combined model also significantly outperformed CTRz measurement in explaining the CDR-sb.

Table 5: **Associations between cognitive scores (CDR-sb, MMSE, MoCA) and petBrain-derived A, $T_2$, and N biomarkers, as well as the combined A/$T_2$/N model, estimated using linear mixed-effects models.** Models were adjusted for age, sex, APOEε4 status, and education.

| | p-value | $R^2$ |
|---|---|---|
| | CDR-sb (N = 718) | |
| **A (CL)** | 9.47e-27 | 0.169 |
| **$T_2$ (CTRz)** | 3.16e-43 | 0.254 |
| **N (HAVAs)** | 3.65e-55 | 0.310 |
| **A/$T_2$/N** | **6.33e-64** | **0.356**[a,b] |

| | MMSE (N = 719) | |
|---|---|---|
| **A (CL)** | 1.13e-30 | 0.190 |
| **$T_2$ (CTRz)** | 1.52e-45 | 0.265 |
| **N (HAVAs)** | 4.03e-45 | 0.263 |
| **A/$T_2$/N** | **2.01e-57** | **0.328[a]** |
| | MoCA (N = 647) | |
| **A (CL)** | 1.56e-34 | 0.231 |
| **$T_2$ (CTRz)** | 1.31e-49 | 0.311 |
| **N (HAVAs)** | 8.62e-51 | 0.316 |
| **A/$T_2$/N** | **6.69e-61** | **0.373[a]** |

[a] Indicates a significant difference (adjusted p-value<0.0125) with model A using a Steiger's test on $R^2$

[b] Indicates a significant difference (adjusted p-value< 0.0125) with model $T_2$ using a Steiger's test on $R^2$

The Steiger's test has been corrected for multiple comparison using Bonferroni correction

## 4. Discussion

In this study, we introduced petBrain, a novel accurate and efficient processing pipeline for amyloid-PET, tau-PET, and structural MRI dedicated to AD research purposes. The petBrain pipeline enables standardized and accessible quantification of amyloid and tau burden using the CL and CTRz scales, ensuring cross-tracer comparability. Additionally, the pipeline provides an age-adjusted measure of brain atrophy through the HAVAs score, derived from lifespan-based brain chart models. After calibration according to the Centiloid and CenTauR project guidelines, we demonstrated that petBrain automatically produces results fully comparable to those obtained with SPM and B-PIP pipelines on ADNI data for amyloid-PET and tau-PET in less than 20 minutes. We further validated petBrain biological relevance by showing correlations with fluid biomarkers and clinical data. The results obtained with petBrain are consistent with the literature, showing significant but imperfect correlations between PET and fluid biomarkers.[32] Indeed, these variables, although strongly associated, reflect different biological processes.[33] Furthermore, the data obtained with petBrain align with the existing literature demonstrating a better association between clinical performance and tau load or neurodegeneration than with amyloid load.[34]

In comparison to existing pipelines, we developed a novel end-to-end solution featuring several key innovations. Notably, the pipeline generates subject-specific ROI masks in under 15 minutes using deep learning-based segmentation of 3D MRI, significantly reducing processing time compared to traditional methods such as FreeSurfer. Additionally, it enables robust biomarker quantification through standardized CL and CTRz scales, ensuring consistency across various amyloid and tau tracers, in contrast to tau SUVr-based methods employed in B-PIP. Furthermore, petBrain is the first pipeline to simultaneously estimate A, $T_2$, and N biomarkers within a unified processing framework contrary to SPM-based pipelines. The web-based implementation removes the need for local computational infrastructure or specialized image processing expertise, making it accessible for broad adoption. Finally, the structured output in the form of a final PDF report (see example in supplementary materials Figure 5) is designed to facilitate the widespread and standardized use of A/$T_2$/N biomarker quantification.

In addition to clinically relevant outputs, such as the binary A+/A-, $T_2$+/$T_2$-, N+/N- status and the quantification of lesion load in CLs and CTRz, petBrain provides advanced information for research purposes. For advanced users, petBrain provides CSV files containing SUVr mean values for 122 brain structures. This information can be utilized, for example, to estimate Braak staging or to detect atypical AD from tau regional SUVr. [35,36,37] Additionally, with the integration of a lifespan model,[20] petBrain offers deviation scores from normal aging for each of the 132 structural volumes derived from T1-weighted MRI, which can be leveraged for differential diagnoses.[38] Experts in image processing can replicate the petBrain pipeline locally by using the freely available version of AssemblyNet (https://github.com/volBrain/AssemblyNet/) along with the calibration equations provided in this article.

Although PVC is not routinely applied in amyloid and tau PET quantification, we chose to incorporate a PVC step into our pipeline. To assess its impact, all experiments were replicated without PVC (see Section 7 of the Supplementary Material). These complementary analyses revealed a high degree of concordance between results obtained with and without PVC, thereby supporting the robustness and reliability of the proposed methodology. While observed differences were minor, a consistent trend favoring the use of PVC was noted.

The limited variations in Centiloid and CenTauRz values are likely attributable to the averaging over large anatomical masks — such as the CenTauR mask, which encompasses over 50,000 voxels — thus mitigating the local influence of partial volume effects. As a result, quantitative estimates remain largely stable across both conditions.

Despite the modest numerical impact, we opted to retain the PVC step due to its enhancement of image quality. This improvement is particularly valuable for the visual assessment of amyloid and tau deposition, offering added benefit in both clinical and research settings (see Supplementary Figure 5 for an illustrative example).

Our pipeline has some limitations. Unlike some amyloid-PET processing pipelines that can be ''MRI-free'', petBrain requires a T1w-MRI. This is also inherent to the ability to obtain the N biomarker with petBrain alongside A and $T_2$ markers. However, it is important to note that petBrain can run without amyloid-PET to provide only $T_2$ and N, or without tau-PET to provide only A and N. In addition, although petBrain provides regional tauopathy progression data by extracting SUVr values from 122 predefined gray matter regions of interest, it currently lacks simplified, clinician-oriented outputs. This limitation is primarily due to threshold variability across different tau tracers. Future developments should aim to incorporate clinically interpretable metrics, such as automatic Braak stage classification based on tau-PET data or assessments of tau-PET positivity in neocortical versus medial temporal regions. Indeed, these indicators are essential for establishing biological staging of AD, as outlined by the AA criteria,[1] and for identifying presymptomatic AD in cognitively unimpaired individuals according to the International Working Group (IWG) criteria.[39] Moreover, future studies should evaluate the generalizability of petBrain across diverse cohorts, including individuals with atypical AD phenotypes, and conduct longitudinal validation to determine its utility in monitoring treatment response to disease-modifying therapies. Finally, we plan to integrate white matter hyperintensities (WMH) in future versions of the pipeline by incorporating the DeepLesionBrain tool[40] for lesion segmentation on FLAIR images. WMH are increasingly recognized as key vascular biomarkers that may interact with Alzheimer's pathology and contribute to cognitive decline.

In conclusion, the petBrain pipeline is a fast and efficient end-to-end pipeline designed for the quantification of A/$T_2$/N biomarkers using PET and MRI. It facilitates the measurement of amyloid and tau loads across various PET tracers using the widely recognized Centiloid and CenTauR scales, ensuring standardized and reliable results. petBrain demonstrates expected concordance with fluid biomarkers and clinical data, making it a robust tool for clinical research applications. The pipeline provides comprehensive report files, along with processed images and segmentation masks, which are easy to interpret and analyze. Additionally, petBrain is freely accessible for research purposes via the user-friendly volBrain platform (www.volbrain.net), ensuring broad accessibility to the scientific community.

# Acknowledgements/Conflicts/Funding Sources/Consent Statement

ACKNOWLEDGMENTS

The ADNI data used in the preparation of this manuscript were obtained from the Alzheimer's Disease Neuroimaging Initiative (ADNI) (National Institutes of Health Grant U01 AG024904). The ADNI is funded by the National Institute on Aging and the National Institute of Biomedical Imaging and Bioengineering and through generous contributions from the following: Abbott, AstraZeneca AB, Bayer Schering Pharma AG, Bristol-Myers Squibb, Eisai Global Clinical Development, Elan Corporation, Genentech, GE Healthcare, GlaxoSmithKline, Innogenetics NV, Johnson & Johnson, Eli Lilly and Co., Medpace, Inc., Merck and Co., Inc., Novartis AG, Pfizer Inc., F. Hoffmann-La Roche, Schering-Plough, Synarc Inc., as well as nonprofit partners, the Alzheimer's Association and Alzheimer's Drug Discovery Foundation, with participation from the U.S. Food and Drug Administration. Private sector contributions to the ADNI are facilitated by the Foundation for the National Institutes of Health (www.fnih.org). The grantee organization is the Northern California Institute for Research and Education, and the study was coordinated by the Alzheimer's Disease Cooperative Study at the University of California, San Diego. ADNI data are disseminated by the Laboratory for NeuroImaging at the University of California, Los Angeles. This research was also supported by NIH grants P30AG010129, K01 AG030514 and the Dana Foundation.

Moreover, this work is based on open access samples. We wish to thank all investigators of these projects who collected these datasets and made them freely accessible.

CONFLICT OF INTEREST STATEMENT

P.C has no conflict of interest to declare. B.M has no conflict of interest to declare. F.M has no conflict of interest to declare. S.M-O has no conflict of interest to declare. J-V. M has no conflict of interest to declare. Independent of this work, NV received research support from Fondation Bettencourt-Schueller, Fondation Servier, Union Nationale pour les Intérêts de la Médecine (UNIM), Fondation Claude Pompidou, Fondation Alzheimer, Banque Publique d'Investissement, Lion’s Club Alzheimer and Fondation pour la Recherche sur l’Alzheimer; travel grant from the Movement Disorders Society, Merz-Pharma, UCB Pharma, and GE Healthcare SAS; is an unpaid local principal investigator or sub-

investigator in NCT05531526 (AR1001, AriBio), NCT06079190 (AL101, GSK), NCT04241068 and NCT05310071 (aducanumab, Biogen), NCT05399888 (BIIB080, Biogen), NCT03352557 (gosuranemab, Biogen), NCT04592341 (gantenerumab, Roche), NCT03887455 (lecanemab, Eisai), NCT03828747 and NCT03289143 (semorinemab, Roche), NCT04619420 (JNJ-63733657, Janssen – Johnson & Johnson), NCT06544616 (JNJ-64042056, Janssen – Johnson & Johnson), NCT04374136 (AL001, Alector), NCT04592874 (AL002, Alector), NCT04867616 (bepranemab, UCB Pharma), NCT04777396 and NCT04777409 (semaglutide, Novo Nordisk), NCT05469360 (NIO752, Novartis), NCT06647498 (remternetug, Washington University School of Medicine); is the French national coordinator in NCT05564169 (VHB937, Novartis); has given unpaid lectures in symposia organized by Eisai and the Servier Foundation; has been an unpaid expert for Janssen – Johnson & Johnson, Eli-Lilly, Novartis. During the past 3 years, VP was a local unpaid investigator or sub-investigator for clinical trials granted by NovoNordisk, Biogen, Janssen and Alector. He received consultant fees for MRI studies in animals from Motac Neuroscience Ltd, outside the submitted work.

FUNDING

This work benefited from the support of the project HoliBrain of the French National Research Agency (ANR-23-CE45-0020-01). Moreover, this project is supported by the Precision and global vascular brain health institute funded by the France 2030 investment plan as part of the IHU3 initiative (ANR-23-IAHU-0001). Finally, this study received financial support from the French government in the framework of the University of Bordeaux's France 2030 program / RRI "IMPACT and the PEPR StratifyAging. This work benefited from the support of the project ChatvolBrain from CNRS. This work was also granted access to the HPC resources of IDRIS under the allocation 2022-AD011013848R1 made by GENCI. This work also benefited from the support of the project PID2023-152127OB-I00 of the Ministerio de Ciencia, Innovacion y Universidades of Spain.

CONSENT STATEMENT

All human subjects of used databases provided informed consent.

LEGENDS

Figure 2 : Global overview of the petBrain pipeline.

Figure 2: Comparison of Centiloid and Centaur values obtained with petBrain and SPM-based pipelines.

Figure 3: Comparison of Centiloid values and Tau SUVr obtained with petBrain and the value provided by B-PIP.

Figure 4: Comparison of Centiloid values and Tau SUVr obtained with SPM-based pipelines and the value provided by B-PIP.

Figure 5: Centiloid (Amyloid), CenTauRz (Tau) and HAVAs (Neurodegeneration) scores on the ADNI dataset (N=821) using petBrain. Amyloid status was established with the B-PIP PET pipeline and ADNI thresholds. The horizontal lines indicate the pathological threshold for each biomarker.

# petBrain: A New Pipeline for Amyloid, Tau Tangles and Neurodegeneration Quantification Using PET and MRI

## *supplementary material*

*Pierrick Coupé[1], Boris Mansencal[1], Floréal Morandat[1], Sergio Morell-Ortega[2], Nicolas Villain[3,4], Jose V. Manjón[2], Vincent Planche[5,6]*

*1 CNRS, Univ. Bordeaux, Bordeaux INP, LABRI, UMR5800, F-33400 Talence, France*

*2 ITACA, Universitat Politècnica de València, 46022 Valencia, Spain*

*3 AP-HP Sorbonne Université, Hôpital Pitié-Salpêtrière, Department of Neurology, Institute of Memory and Alzheimer's Disease, Paris, France*

*4 Institut du Cerveau - ICM, Sorbonne Université, INSERM U1127, CNRS 7225, Paris, France*

*5 CHU Bordeaux, Service de Neurologie des Maladies Neurodégénératives, Centre Mémoire Ressources Recherche, F-33000 Bordeaux, France*

*6 Univ. de Bordeaux, CNRS, UMR 5293, Institut des Maladies Neurodégénératives, F-33000 Bordeaux, France*

## 1. GAAIN Datasets description

In this section, we provide additional information on the GAAIN data used in our study.

- **PiB dataset (Klunk et al. 2015)**: This dataset consists of 79 paired T1-w MRI and PiB PET scans acquired 50–70 minutes post-injection. It includes data from 34 young cognitively normal (yCN) controls and 45 patients with AD. The 34 yCN were under the age of 45 (mean age: 31 ± 6 years; range: 22–43) and were deemed cognitively normal based on a standard neuropsychological and clinical evaluation. The 45 AD patients were diagnosed according to the 1984 NINCDS-ADRDA criteria, had a Clinical Dementia Rating (CDR) Global Score of 0.5 or 1 and were amyloid-positive.

Their amyloid status was based on PiB-positive cutoff method. Their mean age was 67 ± 10 years (range: 50–89).

- **FBP Dataset (Navitsky et al. 2018)**: This dataset consists of 46 paired T1-w MRI scans and PiB PET (50–70 minutes post-injection) and $^{18}$F-Florbetapir PET (50–60 minutes post-injection) scans from 13 yCN and 33 elderly CN. The 13 yCN were under the age of 35 (mean age: 27 ± 4 years; range: 21–35) and had a Mini-Mental State Examination (MMSE) score ranging from 29 to 30. The 33 elderly subjects were further categorized as follows: 17 AD patients (mean age: 67.1 ± 7.1 years; range: 51–76) with MMSE range: 8–26, 7 individuals with MCI (mean age: 80 ± 8 years; range: 64–89) with MMSE range: 25–28, 3 at-risk elderly subjects (mean age: 80 ± 3 years; range: 78–83) with MMSE range: 28–30 and 6 elderly controls (mean age: 63 ± 8 years; range: 51–75) with MMSE range: 27–30.
- **FTM Dataset (Battle et al. 2018):** This dataset consists of 74 paired T1-w MRI scans and PiB PET (50–70 minutes post-injection) and $^{18}$F-Flutemetamol PET (90–110 minutes post-injection) scans from 24 yCN and 50 elderly subjects. The 24 yCN were under the age of 45 (mean age: 37 ± 5 years; range: 30–45). The 50 elderly subjects were categorized as follows: 20 patients with AD dementia (mean age: 69 ± 10 years; range: 60–82) 20 individuals with amnestic MCI (mean age: 73 ± 7 years; range : 57–83) and 10 older healthy controls (mean age: 57 ± 11 years; range : 47–75).
- **FBB Dataset (Rowe et al. 2017):** This dataset consists of 35 paired T1-w MRI scans and $^{18}$F-Florbetaben PET (90–110 minutes post-injection) and PiB PET (50–70 minutes post-injection) scans from 25 elderly subjects and 10 yCN. The 10 yCN were under the age of 45 (mean age: 33 ± 8 years) and had an MMSE score higher than 28. The 25 elderly subjects were categorized as follows: 6 healthy elderly controls (mean age: 71 ± 8 years, MMSE: 29 ± 1), 9 individuals with MCI (mean age: 72 ± 5 years, MMSE: 28 ± 2), 8 patients with AD (mean age: 69 ± 6 years, MMSE: 23

± 3) and 2 patients with frontotemporal dementia (FTD) (mean age: 74 ± 8 years, MMSE: 23 ± 1).

- **NAV Dataset (Rowe et al. 2016):** This dataset consists of 55 paired T1-weighted MRI scans and PiB PET (50–70 minutes post-injection) and $^{18}$F-NAV4694 PET scans from 10 yCN and 45 CN. The 10 yCN were under the age of 45 (mean age: 33 ± 8 years) and had an MMSE score higher than 28. The 45 elderly subjects were categorized as follows: 25 elderly CN (mean age: 74 ± 8 years, MMSE: 29 ± 1), 10 individuals with MCI (mean age: 75 ± 9 years, MMSE: 27 ± 3), 7 AD patients (mean age: 73 ± 11 years, MMSE: 24 ± 2) and 3 FTD patients (mean age: 68 ± 5 years, MMSE: 27 ± 1).
- **FTP Dataset (Villemagne et al. 2023)**: This dataset contains 100 paired T1-w MRI scans and $^{18}$F-Flortaucipir tau PET images from 50 CN controls and 50 patients with AD. The 50 AD patients were aged 65 and older and had a clinical diagnosis of probable AD with an amnestic phenotype. They also had a positive PiB amyloid PET scan and no additional neurological diagnoses. The 50 CN A- were age- and sex-matched to the AD patient group and had a negative amyloid PET scan.

## 2. Centiloid subject-specific mask

To establish the list of structures to include into the petBrain Centiloid mask, we estimated Cohen's d scores for each structure between the yCN A- and Dementia A+ of the PiB dataset. First, we selected all the structures with a Cohen's d score > 5. Afterwards, we added missing contralateral structures to obtain a symmetrical mask. The final list of selected structures is presented in Supplementary Table 1.

Supplementary Table 2: Cohen's d scores for each structure on the young yCN A- and Dementia A+ of the PiB dataset.

| Structure Name | Cohen's d | Label number |
|---|---|---|
| Right anterior cingulate gyrus | 7.33 | 100 |
| Left anterior cingulate gyrus | 6.81 | 101 |
| Left medial frontal cortex | 6.58 | 141 |
| Left precuneus | 6.40 | 169 |
| Right medial frontal cortex | 6.33 | 140 |
| Right sup. frontal gyrus medial segment | 6.27 | 152 |
| Right Accumbens | 6.06 | 23 |
| Right middle cingulate gyrus | 5.94 | 138 |
| Right sup. temporal gyrus | 5.85 | 200 |
| Right middle temporal gyrus | 5.78 | 154 |
| Left subcallosal area | 5.75 | 187 |
| Left sup. frontal gyrus medial segment | 5.67 | 153 |
| Right precuneus | 5.66 | 168 |
| Right posterior cingulate gyrus | 5.65 | 166 |
| Right middle frontal gyrus | 5.63 | 142 |
| Left posterior cingulate gyrus | 5.58 | 167 |
| Left angular gyrus | 5.57 | 107 |
| Right subcallosal area | 5.56 | 186 |
| Left middle cingulate gyrus | 5.48 | 139 |
| Right anterior insula | 5.41 | 102 |
| Right fusiform gyrus | 5.40 | 122 |
| Right inf. temporal gyrus | 5.33 | 132 |
| Left middle temporal gyrus | 5.32 | 155 |
| Left supramarginal gyrus | 5.31 | 195 |

| Right frontal operculum | 5.23 | 118 |
|---|---|---|
| Left anterior insula | 5.20 | 103 |
| Right supramarginal gyrus | 5.19 | 194 |
| Right triangular inf. frontal gyrus | 5.11 | 204 |
| Right angular gyrus | 5.02 | 106 |
| Left sup. temporal gyrus | 4.91 | 201 |
| Left frontal operculum | 4.90 | 119 |
| Left Accumbens | 4.88 | 30 |
| Left inf. temporal gyrus | 4.82 | 133 |
| Left middle frontal gyrus | 4.80 | 143 |
| Left triangular inf. frontal gyrus | 4.72 | 205 |
| Left fusiform gyrus | 4.71 | 123 |

## 3. PiB calibration

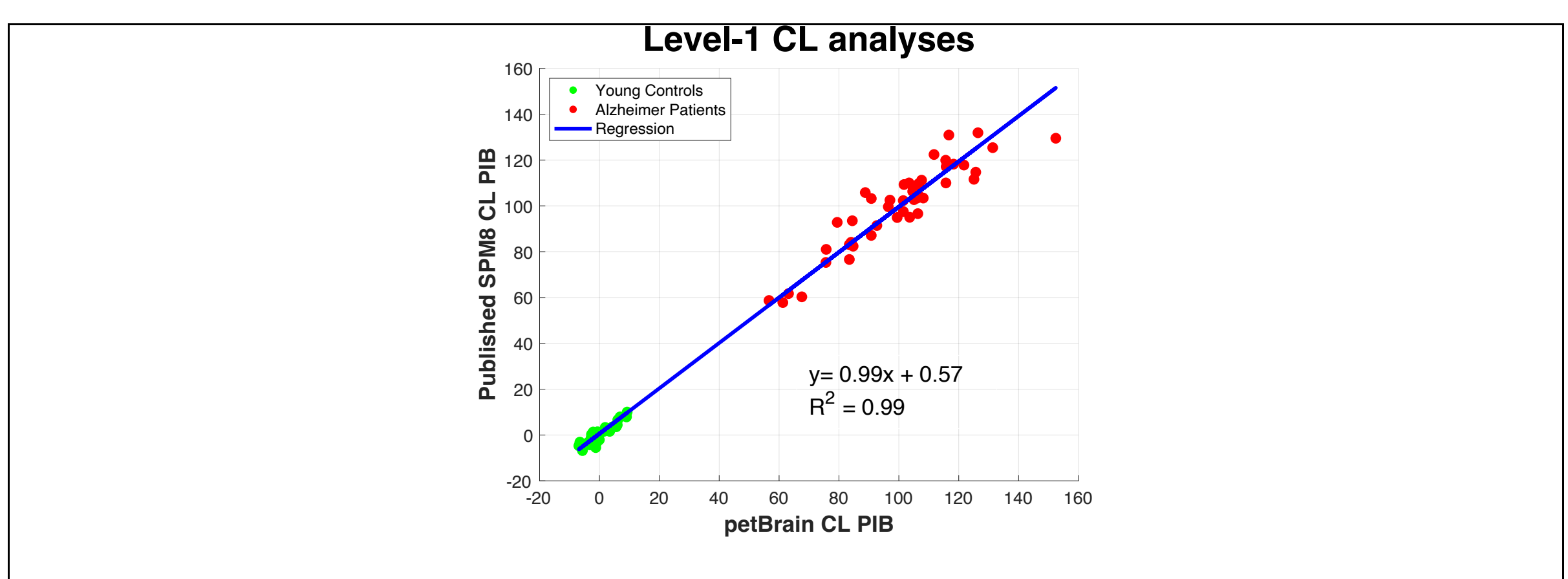

Supplementary Figure 3: **petBrain validation for Centiloid (CL) measure: calibration with the PiB tracer.** The used PiB dataset was composed of 34 young A- cognitively normal subject and 45 A+ patients with dementia.

## 4. Other Amyloid tracers' calibration

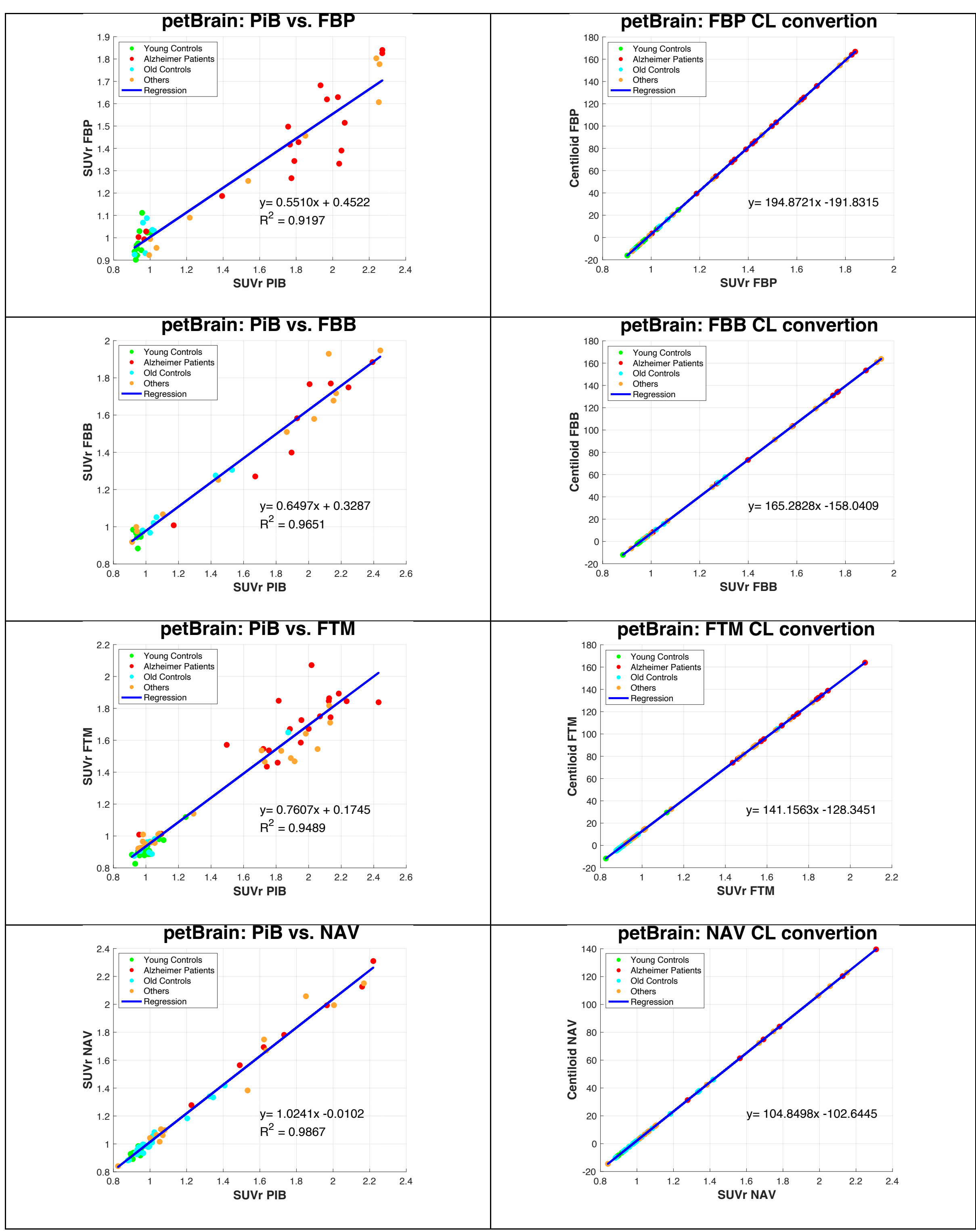

Supplementary Figure 4: Calibration of the FBP, FBB, FTM and NAV amyloid tracers using the corresponding Centiloid Project datasets.

## 5. Centiloid subject-specific mask

For the CenTauR mask, we used the following list of structures – entorhinal area, amygdala, parahippocampal gyrus, fusiform gyrus, inferior and middle temporal gyrus, and temporal pole.

Supplementary Table 3: Cohen's d scores for each structure on the young CN A- and old AD A+ of the FTP dataset.

| Structure Name | Cohen's D | Label number |
|---|---|---|
| Left parahippocampal gyrus | 2.94 | 171 |
| Left entorhinal area | 2.91 | 117 |
| Left Amygdala | 2.65 | 32 |
| Right entorhinal area | 2.63 | 116 |
| Right parahippocampal gyrus | 2.59 | 170 |
| Right Amygdala | 2.54 | 31 |
| Right inf. temporal gyrus | 2.28 | 132 |
| Left inf. temporal gyrus | 2.18 | 133 |
| Right fusiform gyrus | 2.08 | 122 |
| Left middle temporal gyrus | 2.02 | 155 |
| Left fusiform gyrus | 2.01 | 123 |
| Right middle temporal gyrus | 1.99 | 154 |
| Left temporal pole | 1.95 | 203 |
| Right temporal pole | 1.86 | 202 |

Supplementary Table 4: Comparison of the predefined CenTauR masks with our petBrain subject-specific mask on the FTP dataset.

| Mask | $R^2$ |
| --- | --- |
| **Universal** | 0.91 |
| **Mesial-temporal** | 0.90 |
| **Meta-temporal** | 0.98 |
| **Temporo-parietal** | 0.91 |
| **Frontal** | 0.68 |

## 6. CenTauR FTP calibration

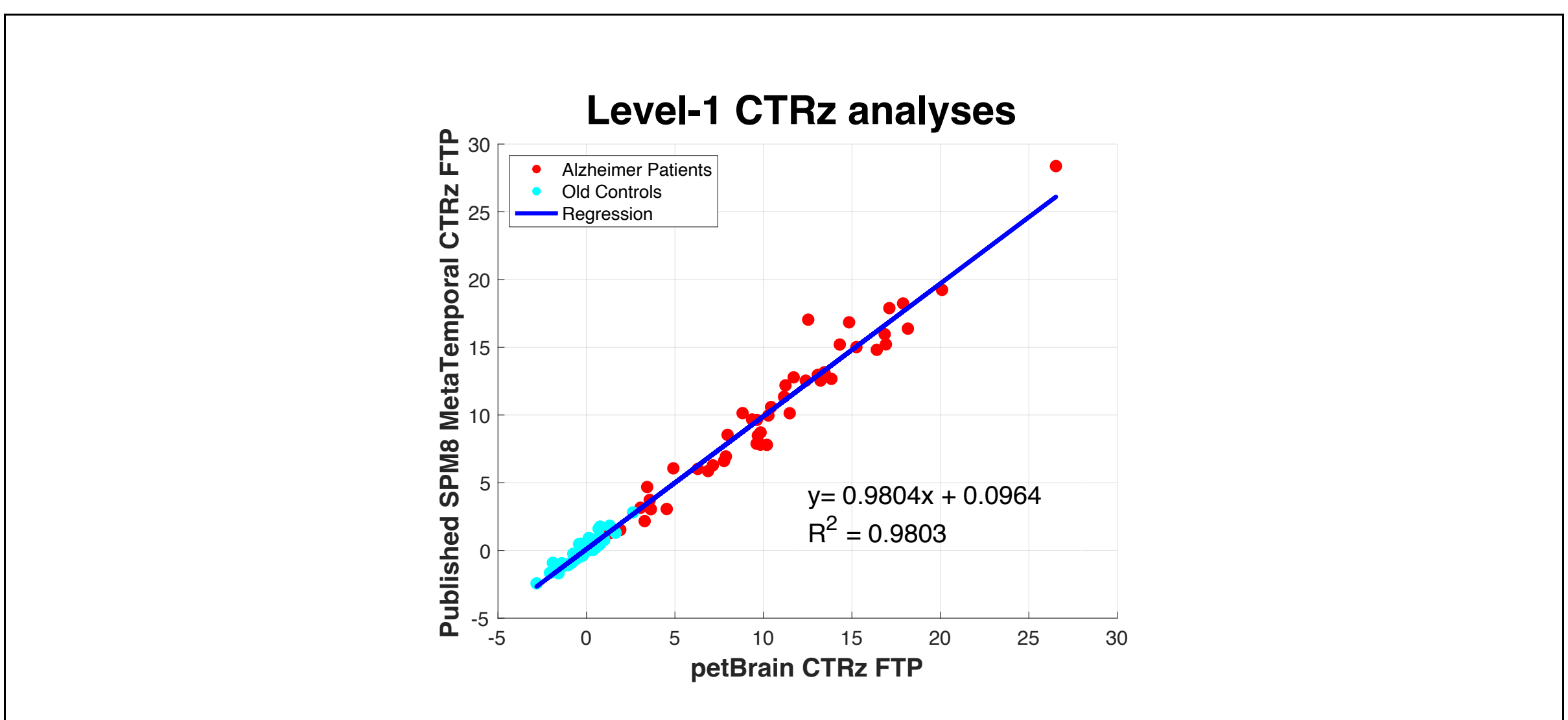

Supplementary Figure 5: **petBrain validation for CenTauRz (CTRz) measure: calibration with FTP tracer**. The used FTP dataset was composed of 50 young A- cognitively normal subject and 50 old A+ patients with AD dementia.

## 7. Validation of PET measurements without Partial Volume Correction (PVC)

In this section, we evaluated the influence of the partial volume correction (PVC) step on the outcomes of our processing pipeline. As illustrated in Supplementary Figure 4, the correlation between petBrain and B-PIP remained highly consistent whether PVC was applied (see Figure 3) or not. Similarly, analyses involving Centiloid and Centaur metrics in relation to AD fluid biomarkers yielded comparable results with and without PVC (Supplementary Table 4). This observation also extended to cognitive scores, for which the associations remained largely unchanged across conditions (Supplementary Table 5). Collectively, these complementary analyses demonstrate a strong concordance between results obtained with and without PVC, supporting the robustness and reliability of the proposed PVC methodology. Although the differences observed were minimal, we noted a consistent trend favoring the application of PVC.

The small variations in Centiloid and CentaurZ values can be attributed to the averaging over large anatomical masks — such as the CenTauR mask, encompassing over 50,000 voxels — which inherently reduces the local impact of partial volume effects. Consequently, quantifications remain largely stable across conditions.

Despite the marginal impact on numerical outcomes, we elected to retain the PVC step in our pipeline due to its contribution to improved image clarity. This enhancement is particularly relevant for the visual assessment of amyloid or tau deposition, providing added value in both clinical and research contexts (see Supplementary Figure 5 for an illustrative example).

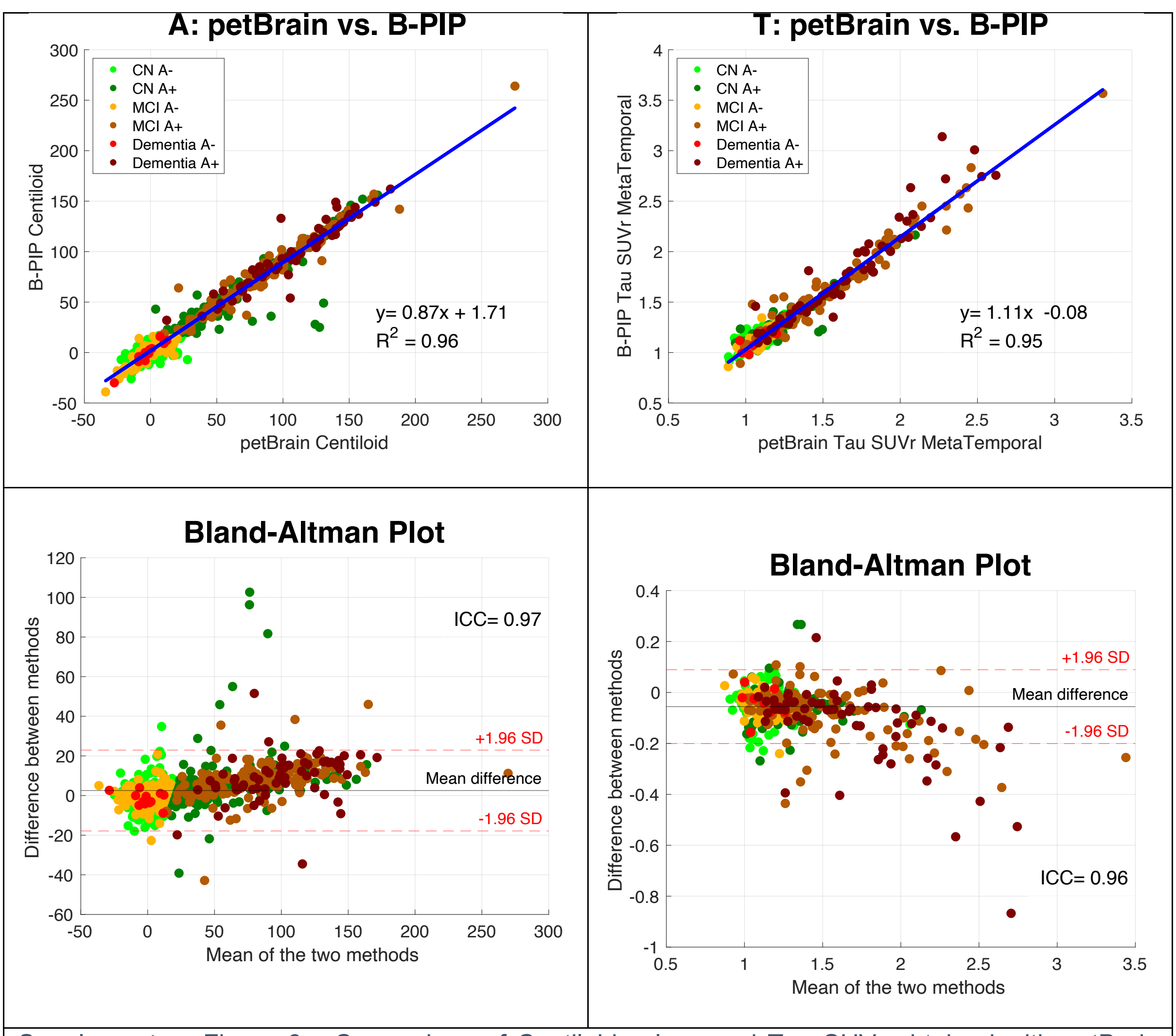

Supplementary Figure 6 : Comparison of Centiloid values and Tau SUVr obtained with petBrain (without PVC step) and the value provided by B-PIP.

Supplementary Table 4: **Association of Centiloid (CL), tau SUVr, CenTauRz (CTRz) produced by petBrain (with and without PVC step) with AD fluid biomarkers.** The linear mixed models were adjusted for age, sex, and APOEε4. The used metrics were p-value of the F-test and $R^2$.

| | p-value | $R^2$ |
|---|---|---|
| **CL ~ CSF Aβ42/40 (N=353)** | | |
| **petBrain** | 2.46e-58 | 0.546 |
| **petBrain wihtout pvc** | 2.59e-58 | 0.545 |
| **CL ~ CSF p-tau (N=350)** | | |
| **petBrain** | 7.1e-27 | 0.311 |
| **petBrain wihtout pvc** | 9.9e-27 | 0.307 |
| **CL ~ log(Plasma p-t217) (N=397)** | | |
| **petBrain** | 2.49e-65 | 0.542 |
| **petBrain wihtout pvc** | 8.28e-64 | 0.534 |
| **Meta-temporal tau SUVr ~ log(Plasma p-t217) (N=397)** | | |
| **petBrain** | 8.23e-54 | 0.475 |
| **petBrain wihtout pvc** | 1.03e-53 | 0.475 |
| **CTRz ~ log(Plasma pT217) (N=397)** | | |
| **petBrain** | 1.48e-56 | 0.492 |
| **petBrain wihtout pvc** | 1.61e-56 | 0.490 |

Supplementary Table 5: **Associations between cognitive scores (CDR-sb, MMSE, MoCA) and petBrain without and with pvc.** Models were adjusted for age, sex, APOEε4 status, and education.

| | p-value | $R^2$ |
|---|---|---|
| | CDR-sb (N = 718) | |
| | wihtout pvc / with pvc | wihtout pvc / with pvc |
| **A (CL)** | 1.52e-26 / 9.47e-27 | 0.168 / 0.169 |
| **$T_2$ (CTRz)** | 6.7e-43 / 3.16e-43 | 0.253 / 0.254 |
| **N (HAVAs)** | 3.65e-55 | 0.310 |
| **A/$T_2$/N** | **7.67e-64 / 6.33e-64** | **0.356[a,b] / 0.356[a,b]** |
| | MMSE (N = 719) | |
| **A (CL)** | 3.54e-30 / 1.13e-30 | 0.185 / 0.190 |
| **$T_2$ (CTRz)** | 3.81e-45 / 1.52e-45 | 0.264 / 0.265 |
| **N (HAVAs)** | 4.03e-45 | 0.263 |
| **A/$T_2$/N** | **2.81e-57 / 2.01e-57** | **0.328[a] / 0.328[a]** |
| | MoCA (N = 647) | |
| **A (CL)** | 2.66e-34 / 1.56e-34 | 0.230 / 0.231 |
| **$T_2$ (CTRz)** | 6.37e-49 / 1.31e-49 | 0.310 / 0.311 |
| **N (HAVAs)** | 8.62e-51 | 0.316 |
| **A/$T_2$/N** | **8e-61 / 6.69e-61** | **0.373[a] / 0.373[a]** |

[a] Indicates a significant difference (adjusted p-value<0.0125) with model A using a Steiger's test on $R^2$

[b] Indicates a significant difference (adjusted p-value< 0.0125) with model $T_2$ using a Steiger's test on $R^2$

The Steiger's test has been corrected for multiple comparison using Bonferroni correction

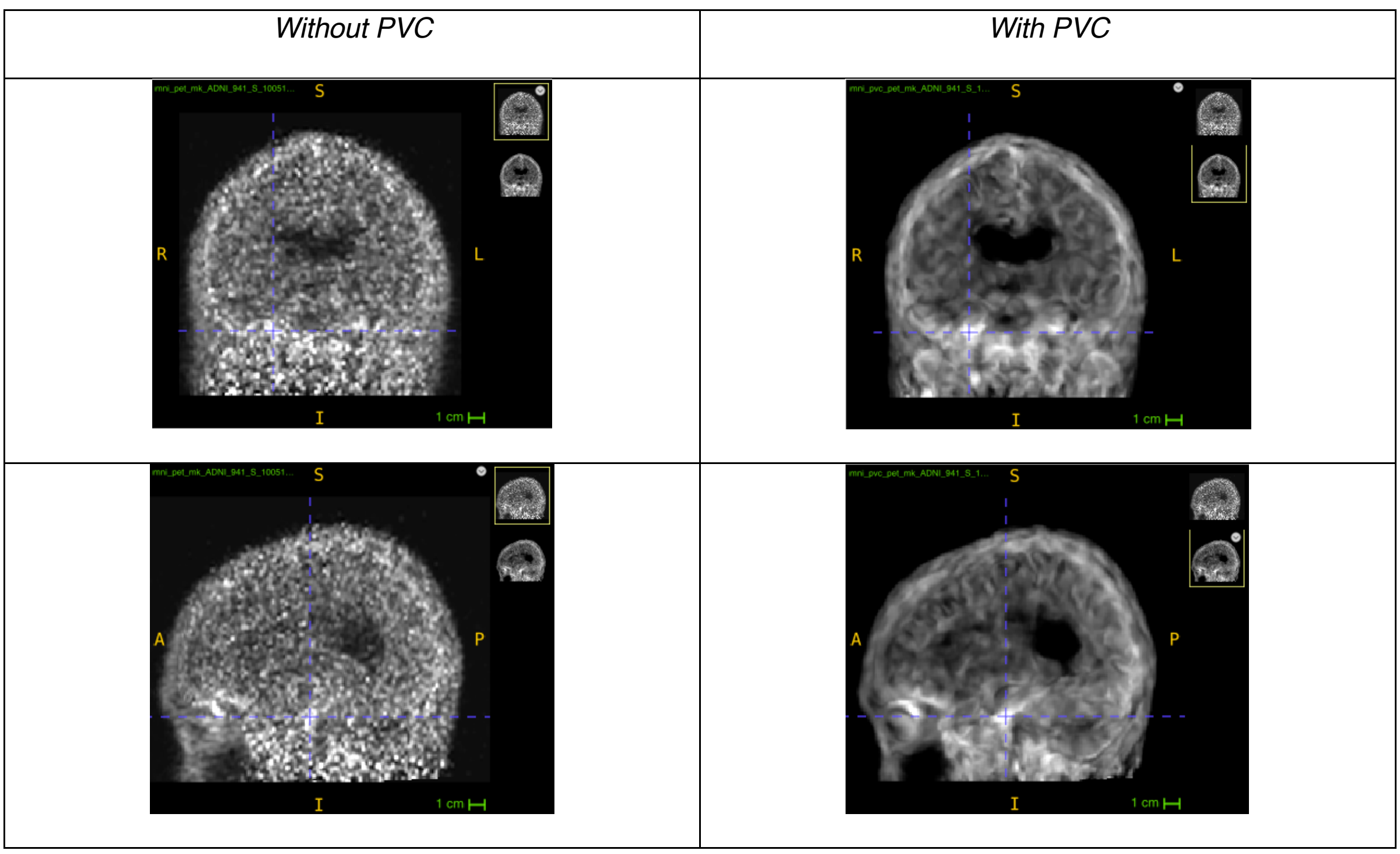

Supplementary Figure 5. Example of Tau PET imaging (MK tracer) in an amyloid-positive subject from the ADNI cohort presenting with clinical symptoms of Alzheimer's disease. The application of PVC enhances the visualization of Tau deposition, particularly in the parahippocampal gyrus, allowing for improved anatomical visualization of tracer uptake.

# 8. Automatically generated report by the web-based VolBrain platform

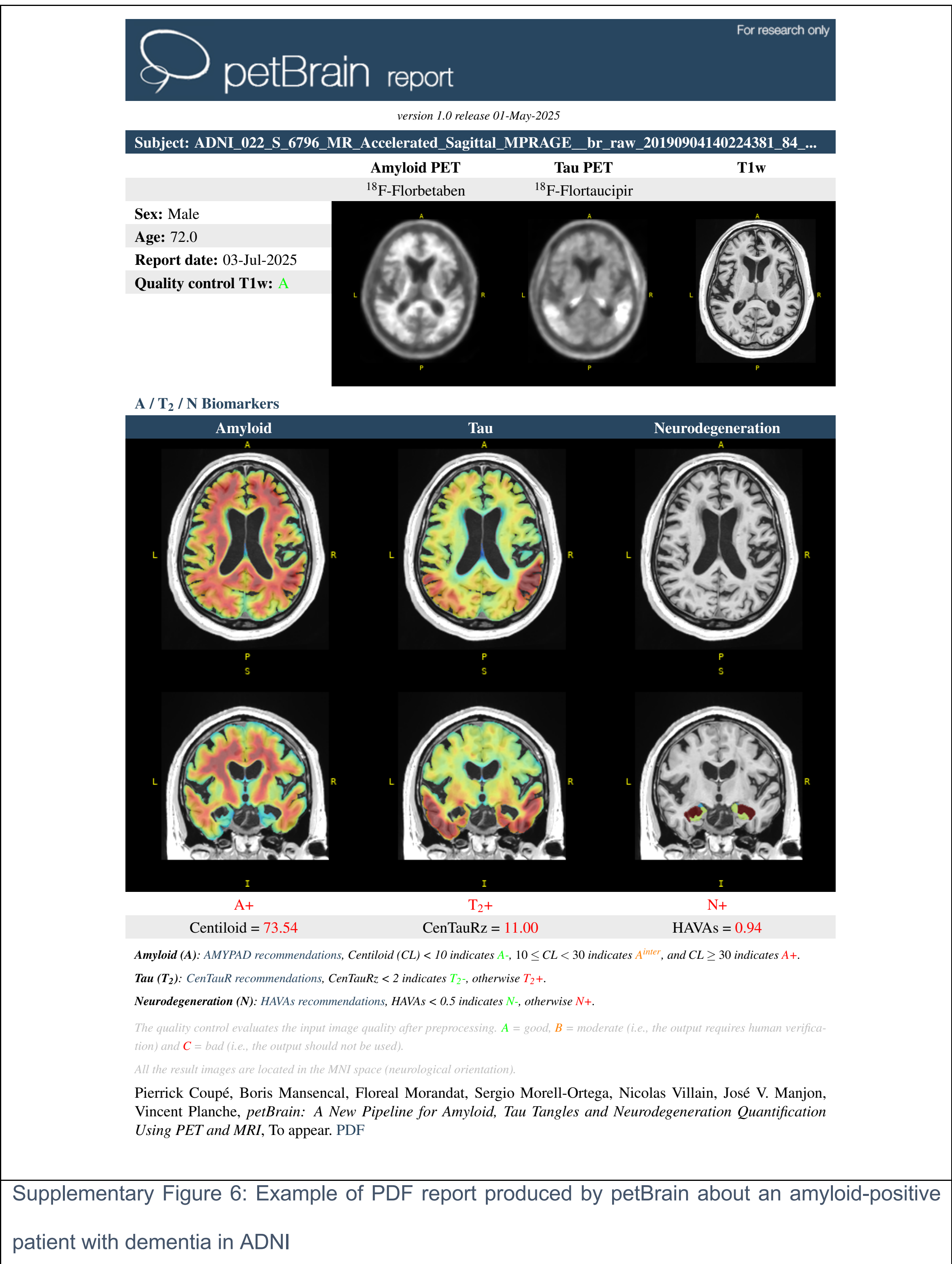

Supplementary Figure 6: Example of PDF report produced by petBrain about an amyloid-positive patient with dementia in ADNI